\documentclass[11pt]{article}\usepackage[]{graphicx}\usepackage[]{color}
\makeatletter
\def\maxwidth{ %
  \ifdim\Gin@nat@width>\linewidth
    \linewidth
  \else
    \Gin@nat@width
  \fi
}
\makeatother

\definecolor{fgcolor}{rgb}{0.345, 0.345, 0.345}

\usepackage{framed}
\makeatletter
 {\par\unskip\endMakeFramed%
 \at@end@of@kframe}
\makeatother

\definecolor{shadecolor}{rgb}{.97, .97, .97}
\definecolor{messagecolor}{rgb}{0, 0, 0}
\definecolor{warningcolor}{rgb}{1, 0, 1}
\definecolor{errorcolor}{rgb}{1, 0, 0}
\newenvironment{knitrout}{}{} 

\usepackage{alltt}

\usepackage[numbers]{natbib}
\usepackage{amsmath}
\usepackage{amssymb}
\usepackage{url}
\usepackage[left=2.5cm,right=2cm, top = 2cm, bottom = 3cm]{geometry}

\newcommand{\poisthin}{\ast}
\IfFileExists{upquote.sty}{\usepackage{upquote}}{}
\begin{document}

\title{A new INARMA(1, 1) model with Poisson marginals}
\author{Johannes Bracher}
\maketitle

\begin{center}
Epidemiology, Biostatistics and Prevention Institute, University of Zurich,\\ Hirschengraben 84, 8001 Zurich, Switzerland\\

\medskip

\texttt{johannes.bracher@uzh.ch}

\bigskip

\noindent\fbox{\parbox{0.8\textwidth}{
\footnotesize
\textbf{Bibliographic note:} This is a pre-print (submitted version before peer review) of a contribution in Steland, A., Rafajlowicz, E., Okhrin, O. (Eds.): \textit{Stochastic Models, Statistics and Their Applications}, p. 323--333, published by Springer Nature Switzerland, 2019. The final authenticated version is available at \url{https://doi.org/10.1007/978-3-030-28665-1_24}.
}}

\end{center}

\bigskip

\abstract{We suggest an INARMA(1, 1) model with Poisson marginals which extends the INAR(1) in a similar way as the INGARCH(1, 1) does for the INARCH(1) model. The new model is equivalent to a binomially thinned INAR(1) process. This allows us to obtain some of its stochastic properties and use inference methods for hidden Markov models. The model is compared to various other models in two case studies.}

\section{Introduction}

Time series of counts are encountered in a broad variety of contexts. Two popular modelling approaches are the INAR (integer-valued autoregressive \cite{Al-Osh1987}) and INGARCH (integer-valued generalized autoregressive conditional heteroscedasticity \cite{Ferland2006}) classes. In this article an extension of the Poisson INAR(1) model is proposed which parallels the generalization of the INARCH(1) to the INGARCH(1, 1) model. We give some properties of the new model, which we refer to as INARMA(1, 1), and point out how to do inference via methods for hidden Markov processes. The performance of the model is compared to various INAR and INGARCH-type models in two case studies.

\section{The Poisson INAR(1) and INARCH(1) models}
\label{sec:inar_inarch}

The Poisson INAR(1) model $\{X_t, t \in \mathbb{Z}\}$ with parameters $\nu > 0$ and $0 < \alpha < 1$ is defined as \cite{Al-Osh1987}
\begin{align}
X_t = I_t + \alpha \circ X_{t - 1}.
\end{align}
The operator $\circ$ denotes binomial thinning, i.e.\ $\alpha \circ Y = \sum_{i = 1}^{Y} Z_i$ with $Z_i \stackrel{\text{iid}}{\sim} \text{Bernoulli}(\alpha)$, implying $\alpha \circ Y \mid Y \sim \text{Bin}(Y, \alpha)$. The sequence $\{I_t, t \in \mathbb{Z}\}$ consists of independent Poisson random variables with rate $\nu$. All thinning operations are performed independently of each other and of $\{I_t\}$. Moreover, at each time $t$, $\alpha \circ X_{t}$ and $I_t$ are independent of all $X_u, u < t$. Marginally the $X_t$ are then Poisson distributed with rate $\nu/(1 - \alpha)$; the autocorrelation function is $\rho(h) = \alpha^h$.

The Poisson INARCH(1) model $\{X_t, t \in \mathbb{Z}\}$ is usually defined as \cite{Weiss2010}
\begin{align}
X_t \mid X_{t - 1}, X_{t - 2}, \dots & \sim \text{Pois}(\lambda_t)\label{eq:x_inarch}; \ \ \lambda_t = \nu + \alpha X_{t - 1}
\end{align}
with $\nu > 0, \alpha \geq 0$, but can also be formulated as (compare \cite{Weiss2015})
\begin{equation}
X_t = I_t + \alpha \poisthin X_{t - 1}\label{eq:ingarch_thinning}
\end{equation}
where $\{I_t\}$ is again a sequence of independent Poisson random variables with rate $\nu$. We define the operator $\poisthin$ as $\alpha\poisthin Y \mid Y \sim \text{Pois}(\alpha Y)$ where for $\alpha Y = 0$ we include the degenerate Poisson distribution $\text{Pr}(\alpha\poisthin Y = 0) = 1$. Note that while $X_{t - 1}$ in \eqref{eq:ingarch_thinning} is integer-valued, $\alpha \poisthin Y$ is also defined for real-valued $Y \geq 0$. If $\alpha < 1$, the process $\{X_t\}$ is stationary with $\mathbb{E}(X_t) = \nu/(1 - \alpha)$ and $\text{Var}(X_t) = \mathbb{E}(X_t)/(1 - \alpha^2)$, i.e.\ $X_t$ is over-dispersed for $\alpha > 0$. The autocorrelation function is again $\rho(h) = \alpha^h$.

\section{Extension to models with ARMA(1, 1)-like covariance structure}
\label{sec:inarma_ingarch}

The INARCH(1) model can be extended to the Poisson INGARCH(1, 1) model \cite{Ferland2006}
\begin{align}
X_t \mid X_{t - 1}, X_{t - 2}, \dots & \sim \text{Pois}(\lambda_t)\label{eq:ingarch1}\\
\lambda_t & = \nu + \alpha X_{t - 1} + \beta \lambda_{t - 1}\label{eq:ingarch2}
\end{align}
with $\nu > 0$ and $\alpha, \beta \geq 0$. In the following we assume $\{X_t\}$ to be stationary, which is the case if $\alpha + \beta < 1$. We can then express it using the operator $\poisthin$ from \eqref{eq:ingarch_thinning}. Consider
\begin{equation*}
S_{t} = \frac{\lambda_t - \frac{\nu}{1 - \beta}}{1 - \beta}
\end{equation*}
which after some simple algebra leads to
\begin{align*}
\lambda_t & = (1 - \beta) S_t + \frac{\nu}{1 - \beta}\\
S_t & = \beta S_{t - 1} + \frac{\alpha}{1 - \beta}\cdot X_{t - 1}\ .
\end{align*}
Note that the recursive definition \eqref{eq:ingarch2} of $\lambda_t$ implies $\lambda_t \geq \sum_{d = 0}^\infty \nu\beta^d = \nu/(1 - \beta)$ so that non-negativity of $S_t$ is ensured. An alternative display of \eqref{eq:ingarch1}--\eqref{eq:ingarch2} is then
\begin{align}
X_t & = \phi \poisthin S_t + I_t\label{eq:x_t_ingarch}\\
S_{t} & = (1 - \phi) S_{t - 1} + \kappa X_{t - 1}\label{eq:s_t_ingarch}
\end{align}
with $I_t \stackrel{\text{iid}}{\sim} \text{Pois}(\tau)$ and
\begin{align*}
\tau = \frac{\nu}{1 - \beta}; \ \ \ \phi = 1 - \beta; \ \ \ \kappa & = \frac{\alpha}{1 - \beta} \ .
\end{align*}
Stationarity of $\{X_t\}$ implies $0 \leq \kappa < 1$ and one obtains \cite{Ferland2006}
\begin{align*}
\mathbb{E}(X_t) & = \frac{\tau}{1 - \kappa}; \ \ \
\text{Var}(X_t) = \frac{1 - \xi^2 + \kappa^2\phi^2}{1 - \xi^2}\cdot \mathbb{E}(X_t)\\
\rho(h) & =  \frac{1 - \xi^2 + \kappa^2\phi^2 + \kappa\phi(1 - \phi)}{1 - \xi^2 + \kappa^2\phi^2}\cdot \kappa\phi\xi^{h - 1}
\end{align*}
with $\xi = 1 - \phi(1 - \kappa)$, i.e.\ the second-order properties of an ARMA(1, 1) process.

We now suggest a similar generalization of the Poisson INAR(1) model which we call Poisson INARMA(1, 1). It is defined as $\{X_t, t \in \mathbb{Z}\}$ with
\begin{align}
X_t & = \phi \circ S_t + I_t\label{eq:X_inarma}\\
S_t & = S_{t - 1} - (X_{t - 1} - I_{t - 1}) + \kappa \circ X_{t - 1}\label{eq:S_inarma}
\end{align}
where $I_t \stackrel{\text{iid}}{\sim} \text{Pois}(\tau), \tau > 0$ and $0 < \phi \leq 1,  0 < \kappa < 1$. Again, all thinning operations are independent of each other and of $\{I_t\}$. At each $t$, $\phi \circ S_{t}$, $\kappa\circ X_{t}$ and $I_t$ are independent of all $X_u, S_u, u < t$ and, given $X_t$, $\kappa\circ X_{t}$ is independent of $S_t$. This formulation parallels \eqref{eq:x_t_ingarch}--\eqref{eq:s_t_ingarch} as, using $X_{t - 1} - I_{t - 1} = \phi \circ S_{t - 1}$, it is easily seen that
$$
S_t \overset{d}{=} (1 - \phi) \circ S_{t - 1} + \kappa\circ X_{t - 1}\label{eq:S_inarma_b}.
$$
However, \eqref{eq:S_inarma} implies a dependence between the two thinnings $\phi\circ S_t$ and $(1 - \phi)\circ S_t$, entering into $X_t$ and $S_{t + 1}$, respectively, as they are forced to sum up to $S_t$. Unlike in the INGARCH model\footnote{The INGARCH(1, 1) model, too, can be expressed with a discrete-valued process $\{S_t\}$, just set $S_t = S_{t - 1} - (X_{t - 1}  - I_{t - 1}) + \kappa \poisthin X_{t - 1}$ in \eqref{eq:X_inarma}--\eqref{eq:S_inarma}. Details are omitted due to space constraints.} \eqref{eq:x_t_ingarch}--\eqref{eq:s_t_ingarch}, $S_t$ is discrete-valued here (it can be shown to be an INAR(1) process with $S_t = J_t + \xi\circ S_{t - 1}; J_t \sim \text{Pois}(\kappa\tau)$). This is necessary to ensure well-definedness of $\phi \circ S_t$ and achieved by replacing the multiplications from \eqref{eq:s_t_ingarch} by binomial thinnings.

As in an INAR(1) model, the $X_t$ are marginally Poisson distributed under model \eqref{eq:X_inarma}--\eqref{eq:S_inarma}, the rate being
\begin{equation}
\mathbb{E}(X_t) = \text{Var}(X_t) = \tau/(1 - \kappa).\label{eq:mean_inarma}
\end{equation}
The autocorrelation function is
\begin{equation}
\rho(h) = \phi\kappa\xi^{h - 1}\label{eq:corr_inarma}
\end{equation}
where again $\xi = 1 - \phi(1 - \kappa)$. These properties are easy to show using the representation of $\{X_t\}$ as a binomially thinned INAR(1) process, see next section. Thus the new model, too, has the second-order properties\footnote{As mentioned in \cite{Ferland2006}, Lemma 2, the fact that the autocovariance structure of $\{X_t\}$ coincides with that of a stationary ARMA(1, 1) process is sufficient for $\{X_t\}$ to be an ARMA(1, 1) process itself.} of an ARMA(1, 1) process, justifying the name INARMA(1, 1). Note, however, that the formulation differs from other models referred to as INARMA in the literature (e.g.\ \cite{Neal2006}). The INAR(1) model corresponds to the boundary case $\phi = 1$ of the new class. In comparison to the INGARCH(1, 1) model with the same parameters the new model has lower dispersion and its autocorrelation function is damped if $\phi < 1$.

\section{Alternative displays of INARMA(1, 1) and link to other models}

\begin{figure}[h]
\center
\includegraphics[scale = 1.1]{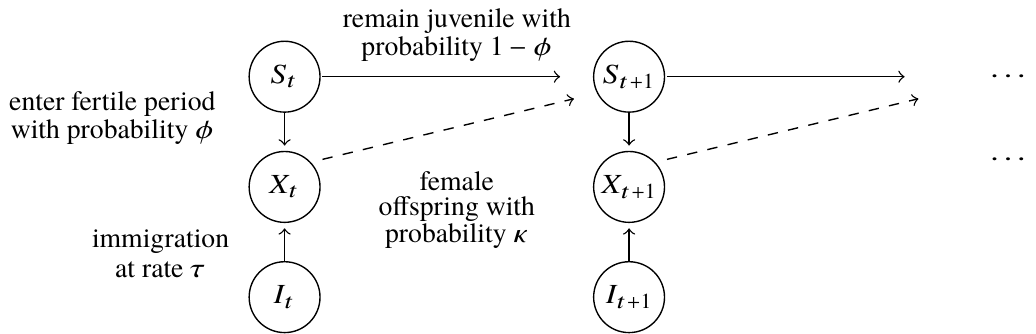}
\caption{\small Interpretation of the INARMA(1, 1) process in the form of a flow diagram.}
\label{fig:interpretation}
\end{figure}

The INARMA(1, 1) model can be interpreted as follows: $X_t$ is the number of fertile females in a population and $S_t$ is the (unobserved) number of juvenile,  i.e.\ not yet fertile females. $I_t$ is the number of fertile female immigrants (there is no immigration of juveniles). Females do not die before reaching fertility and at each time of their juvenile period have a probability of $\phi$ to transition to the fertile state. They stay fertile for exactly one time period and can have at most one female offspring, the probability of which is $\kappa$. A graphical display of such a system can be found in Figure \ref{fig:interpretation}.

\noindent The time from a female's birth to its fertile period obviously follows a geometric distribution with parameter $\phi$. We can use this to express the model as an INAR($\infty$) model
\begin{align}
X_t & = \sum_{i = 1}^\infty \alpha_i \circ X_{t - i} \ \ + \ \ I_t; \ \ \ I_t \stackrel{\text{iid}}{\sim} \text{Pois}(\tau) \label{eq:inar_inf}
\end{align}
with $\alpha_i = \kappa\phi(1 - \phi)^{i - 1}, i = 1, 2, \dots$ and dependent thinning operations given by
\begin{align}
B_t \ \mid X_t & \sim \text{Bin}(X_t, \kappa)\label{eq:B_t}\\
A_{t}^{(j)} & \stackrel{\text{iid}}{\sim} \text{Geom}(\phi), \ \ \ j = 1, \dots B_{t}\label{eq:At}\\
\alpha_i \circ X_{t} & =\sum_{j = 1}^{B_{t}} I(A^{(j)}_{t} = i), \ \ \ i = 1, 2,\dots \label{eq:coupling_thinning_geom}
\end{align}
Here, $B_t$ is the number of female offspring born in $t$ and  $A_t^{(j)}$ is the waiting time until fertility for the $j$th of the females born at time $t$. The definition \eqref{eq:B_t}--\eqref{eq:coupling_thinning_geom} of the dependent thinnings implies that $\alpha_i \circ X_t \mid X_t \sim \text{Bin}(X_t, \alpha_i)$ for $i = 1, 2, \dots$ under the constraint $\sum_{i = 1}^\infty\alpha_{i} \circ X_t \leq X_t$.

The representation \eqref{eq:inar_inf}--\eqref{eq:coupling_thinning_geom} nicely illustrates the relationship of the INARMA(1, 1) model to other common models. Replacing the geometric waiting time distribution in \eqref{eq:At} by a one-point distribution with $\text{Pr}(A_t^{(j)} = 1) = 1$ yields the INAR(1) model while a categorical distribution with support $1, \dots, p$ gives the INAR($p$) model by Alzaid and Al-Osh \cite{Alzaid1990} (see \cite{Bracher2019} for details). Replacing the binomial offspring distribution in \eqref{eq:B_t} by a Poisson distribution, i.e.\ setting
$$
B_t \ \mid X_t \sim \text{Pois}(\kappa X_t)\label{eq:B_t_Pois}
$$
yields the INGARCH(1, 1) model. Due to space restrictions, we do not detail on this, but it is straightforward to show using the INARCH($\infty$) representation of the INGARCH(1, 1) model (\cite{Weiss2018}, p.76) and some basic properties of the Poisson distribution. The INARCH(1) and INARCH($p$) models can be obtained by using again one-point and categorical waiting time distributions in \eqref{eq:At}.

We recently encountered INAR($\infty$) models of type \eqref{eq:inar_inf}--\eqref{eq:coupling_thinning_geom} in \cite{Bracher2019} where we extended work by Fern\'andez-Fontelo et al \cite{Fernandez-Fontelo2016} on underreported INAR models. We showed that $\{X_t, t \in \mathbb{Z}\}$ is equivalent to a binomially thinned INAR(1) model $\{\tilde{Y}_t, t \in \mathbb{Z}\}$ given by
\begin{align}
Y_t & = J_t + \xi \circ Y_{t - 1} \label{eq:equiv_INAR1_Y}\\
\tilde{Y}_t \mid Y_t & \sim \text{Bin}(Y_t, \phi\kappa/\xi)\label{eq:reparam_INAR1_Ytilde}
\end{align}
with $J_t \stackrel{\text{iid}}{\sim} \text{Pois}(\tau\xi/\kappa)$ and, as before, $\xi = 1 - \phi(1 - \kappa)$. This represents an interesting parallel to the Gaussian ARMA(1, 1) model which, as shown for instance in \cite{Staudenmayer2005}, can be obtained by adding homoscedastic measurement error to a Gaussian AR(1) process. This third representation of the process makes the derivation of equations \eqref{eq:mean_inarma}--\eqref{eq:corr_inarma} easy (see \cite{Fernandez-Fontelo2016}, Section 2 and Appendix A; our model corresponds to the special case $\omega = 1$ of the class discussed there). Also, it implies that many properties of the INAR(1) process translate to the INARMA(1, 1) model, e.g.\ the marginal Poisson distribution and time-reversibility \cite{Schweer2015}.

\section{Inference}

Inference for higher-order INAR models with dependent thinning operations is challenging as the likelihood is generally intractable \cite{Alzaid1990}. For our model, however, we can exploit the representation \eqref{eq:equiv_INAR1_Y}--\eqref{eq:reparam_INAR1_Ytilde} as a binomially thinned INAR(1) process. As described in Fern\'andez-Fontelo et al \cite{Fernandez-Fontelo2016}, Section 3.2, the forward algorithm \cite{Zucchini2009}, a standard method for inference in hidden Markov models, can be applied to evaluate the likelihood of this model (again note that our model corresponds to the special case $\omega = 1$ of the class treated in \cite{Fernandez-Fontelo2016}). As the state space of the latent process $\{Y_t\}$ is infinite, truncation at some reasonably large value $Y^{\max}$ is required. The maximum of the log-likelihood is then obtained by numerical optimization.

\section{Case studies}

\begin{figure}[htb]
\begin{knitrout}
\definecolor{shadecolor}{rgb}{0.969, 0.969, 0.969}\color{fgcolor}
\center
\includegraphics[width=0.9\maxwidth]{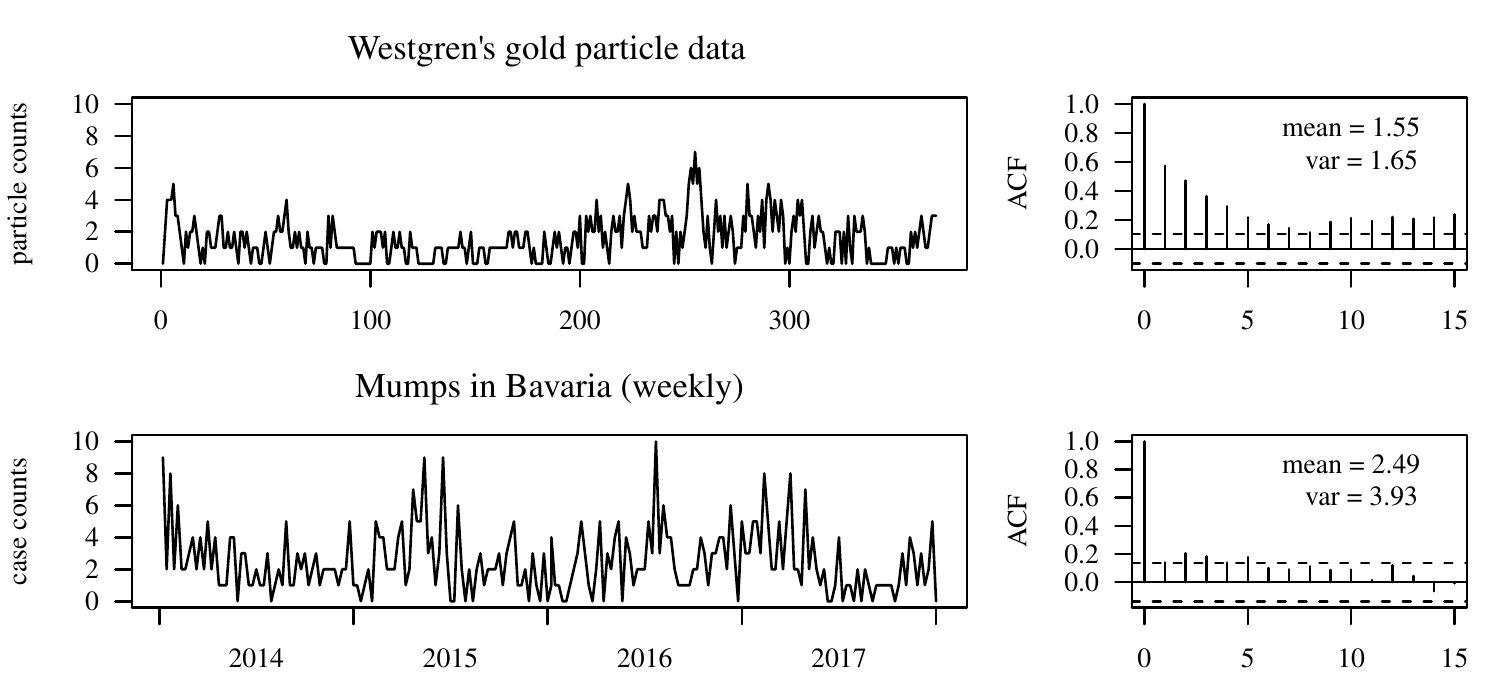} 

\end{knitrout}
\label{fig:data}
\caption{\small Case studies: gold particle counts and weekly numbers of reported
 mumps cases in Bavaria.}
\end{figure}

We apply the four models from Sections \ref{sec:inar_inarch} and \ref{sec:inarma_ingarch} to two data sets. The first example consists of the gold particle counts from Westgren \cite{Westgren1916}, a data set which is often used in the literature. For instance Wei{\ss} (\cite{Weiss2018}, p.48) applies Poisson INAR(1), INAR(2) and CINAR(2) models to these data. To make our results comparable to these analyses we fit all models to observations 501--870 of Series (C). As a second example we use weekly counts of mumps in Bavaria, Germany, from week 1/2014 to week 52/2017 (downloaded from \texttt{www.survstat.rki.de} on 8 Oct 2018). Mumps, a viral disease, used to be a common childhood disease. Since the introduction of a vaccine in the 1950s it has become rare in Germany, but remains under routine surveillance. The data are displayed in Figure \ref{fig:data}. Both time series show slowly decaying autocorrelation functions, indicating that relaxing the AR(1) assumption may be beneficial. While the particle counts are approximately equidispersed (mean 1.55, variance 1.65) the mumps data show some overdispersion (mean 2.49, variance 3.93).

Table \ref{tab:gold} shows parameter estimates and AIC values for the gold particle data. For comparison we added the results Wei{\ss} \cite{Weiss2018} obtains for INAR(2) and CINAR(2) models (see there for details). To assess the out-of-sample predictive performance we computed mean log scores ($\overline{\text{logS}}$, \cite{Czado2009}) of one-step-ahead forecasts for the second half of the time series. For each of these forecasts the models were re-fitted to the data which were already available at the respective time point (``rolling forecast''). Note that the log score is negatively oriented, i.e.\ smaller values are better.

\begin{table}[htb]
\caption{\small Parameter estimates, AIC values and mean log scores for the gold particle data. Mean log scores for INAR(2) and CINAR(2) were computed using code from the supplement of \cite{Weiss2018}.}
\label{tab:gold}
\center
\footnotesize
\begin{tabular}{p{4cm} p{1cm} p{1cm} p{1cm} p{1cm} p{1.5cm} p{1.5cm}}
\hline\noalign{\smallskip}
Model & \multicolumn{4}{c}{Parameter} & AIC & $\overline{\text{logS}}$\\
\noalign{\smallskip}\hline\noalign{\smallskip}
& $\hat\nu$ & $\hat\alpha$ & $\hat\alpha_2$ & $\hat{\lambda}_1$ & \\
\noalign{\smallskip}\hline\noalign{\smallskip}
Poisson INAR(1) &
 0.73 & 0.53 &  &  & 1040 & 1.642 \\ 
  
Poisson INAR(2) & 0.54 & 0.47 & 0.18 & & 1027 & 1.610\\
Poisson CINAR(2) & 0.60 & 0.41 & 0.19 & & 1027 & 1.611\\
Poisson INARCH(1) &
 0.75 & 0.52 &  & 0.00 & 1057 & 1.624 \\ 
  
\noalign{\smallskip}\hline\noalign{\smallskip}
& $\hat\tau$ & $\hat\phi$ & $\hat\kappa$ & $\hat S_1$ & AIC\\
\noalign{\smallskip}\hline\noalign{\smallskip}
Poisson INARMA(1, 1) &
 0.31 & 0.67 & 0.80 &  & 1014 & 1.577 \\ 
  
Poisson INGARCH(1, 1) &
 0.47 & 0.54 & 0.70 & 1.85 & 1047 & 1.592 \\ 
  
\noalign{\smallskip}\hline\noalign{\smallskip}
\end{tabular}
\end{table}

\noindent The INARMA(1, 1) model has the best in-sample and out-of-sample performance. Interestingly it also outperforms the two AR(2) models from \cite{Weiss2018}, indicating that observations more than two time steps back still contain additional information.

The corresponding results for the mumps data can be found in Table \ref{tab:mumps}. While the INARMA(1, 1) model again represents a considerable improvement compared to the INAR(1), the INGARCH(1, 1) model performs best. This is not surprising given the overdispersion in the data.

\begin{table}[h!]
\caption{\small Parameter estimates, AIC values and mean log scores for the mumps data.}
\label{tab:mumps}
\center
\footnotesize
\begin{tabular}{p{4cm} p{1cm} p{1cm} p{1cm} p{1cm} p{1.5cm} p{1.5cm}}
\hline\noalign{\smallskip}
Model & \multicolumn{4}{c}{Parameter} & AIC & $\overline{\text{logS}}$\\
\noalign{\smallskip}\hline\noalign{\smallskip}
& $\hat\nu$ & $\hat\alpha$ & & $\hat{\lambda}_1$ & \\
\noalign{\smallskip}\hline\noalign{\smallskip}
Poisson INAR(1) &
 2.21 & 0.11 &  &  & 842 & 2.017 \\ 
  
Poisson INARCH(1) &
 2.08 & 0.15 &  & 9.01 & 832 & 2.010 \\ 
  
\noalign{\smallskip}\hline\noalign{\smallskip}
& $\hat\tau$ & $\hat\phi$ & $\hat\kappa$ & $\hat S_1$ & AIC\\
\noalign{\smallskip}\hline\noalign{\smallskip}
Poisson INARMA(1, 1) &
 1.21 & 0.25 & 0.52 &  & 827 & 1.963 \\ 
  
Poisson INGARCH(1, 1) &
 1.12 & 0.26 & 0.52 & 18.97 & 816 & 1.955 \\ 
  
\noalign{\smallskip}\hline\noalign{\smallskip}
\end{tabular}
\end{table}

\section{Discussion}

We suggested an INARMA(1, 1) model with Poisson marginals which draws conceptually from the INGARCH(1, 1) model \cite{Ferland2006} and the INAR($p$) model by Al-Osh and Alzaid \cite{Alzaid1990}. We provided an alternative representation in terms of an offspring distribution and a waiting time distribution which enlightens the close relation to several existing models from the literature. A third representation as a binomially thinned INAR(1) process turned out to be useful for inference and to obtain some stochastic properties. In our case studies the model performed favourably for equidispersed data. For overdispersed data it outperformed the INAR(1) model, but achieved lower performance than the INGARCH(1, 1) model. This raises the question how overdispersion could be accommodated in the model. Alternative immigration distributions could be considered, but the model would then no longer have a representation as a binomially thinned version of a Markov process. Obtaining its stochastic properties and evaluating the likelihood would get more involved. Another open question is how higher-order INARMA($p, q$) models could be defined.

\bigskip

\noindent \textbf{Data and code} Data and \texttt{R} code are available at \texttt{\url{github.com/jbracher/inarma}.}

\bigskip

\noindent \textbf{Acknowledgements} The author thanks Christian H. Wei{\ss} for helpful feedback.

\bibliographystyle{plain}
\bibliography{bib_inarma}

\end{document}